\documentclass[10pt,conference,letterpaper,onecolumn]{IEEEtran}
\usepackage{times,amsmath,epsfig,ulem}
\usepackage{color}
\usepackage{amsthm}
\usepackage{amssymb}

\def\ie{\textit{i.e., }}

\title{ENHANCED BLIND DECODING of TARDOS CODES\\
 with NEW MAP-BASED FUNCTIONS}	
\author{
{Mathieu Desoubeaux{\small $~^{\#1}$}, C\'edric Herzet{\small $~^{*2}$}, William Puech{\small $~^{\#}$}, Ga\"etan Le Guelvouit{\small $~^{o}$}}%
\vspace{1.6mm}\\
\fontsize{10}{10}\selectfont\itshape
$^{\#}$\,University of Montpellier 2 - LIRMM - 34095 - Montpellier Cedex 5 - France\\

$^{*}$\,INRIA - Centre Rennes - Bretagne Atlantique - Equipe Fluminance - 35000 - Rennes - France\\
$^{o}$\,Orange Labs - 35512 - Cesson-S\'evign\'e - France

\\
\fontsize{9}{9}\selectfont\ttfamily\upshape
$^{1}$\,mathieu.desoubeaux@lirmm.fr\\
$^{2}$\,cedric.herzet@inria.fr%
}

\begin{document}
\maketitle

\begin{abstract}	
This paper presents a new decoder for probabilistic binary traitor tracing codes under the marking assumption. 
It is based on a binary hypothesis testing rule which integrates a collusion channel relaxation so as to obtain numerical and simple accusation functions. This decoder is blind as no estimation of the collusion channel prior to the accusation is required. Experimentations show that using the proposed decoder gives better performance than the well-known symmetric version of the Tardos decoder for common attack channels. 
\end{abstract}


\section{Introduction}

Active fingerprinting, also known as traitor tracing, first introduced in \cite{wagner_fingerprinting_1983}, aims at finding the leak of an illegal redistribution of copyrighted digital contents. This goal requires to personalize each delivered content by embedding a sequence of symbols associated to each user. 

%

Recent trends to generate such sequences focus on probabilistic codes since they allow for low error probabilities (namely the sum of the false alarm probability and the false negative probability) with 
affordable code lengths and small alphabet's sizes. The performance of the probabilistic codes is usually measured in terms of the minimum code length required to achieve a given error probability. 

One of the most efficient probabilistic codes have been proposed by Tardos in ~\cite{tardos_optimal_2008}. These codes rely on the so-called ``marking assumption", first introduced by Boneh and Shaw in~\cite{boneh_collusion-secure_1998}. 
In particular, the Tardos code was the first one proposed in the literature whose length, say $m$, scales as $\mathcal{O}(c^2 \ln(\epsilon_1^{-1}))$, where $\epsilon_1$ represents the false alarm probability and $c$ is the maximum number of colluders. Tardos code gets therefore very close to the lower bound on the code length proved in ~\cite{peikert_lower_2003} and ~\cite{tardos_optimal_2008}, which states that $m = \Omega(c^2 \ln(\epsilon_1^{-1}))$ for random codes and any number of users $n$ with $n \geq c+1$ users. 

Since Tardos seminal work, many efforts have been devoted to further improve the efficiency and the effectiveness of his code. First, the authors of~\cite{blayer_improved_2008, skoric_tardos_2008} aimed at reducing the constant appearing in the code-length bound. In \cite{nuida_improvement_2007, kuribayashi_systematic_2008}, the authors respectively focussed on the improvement of the memory consumption and the decoding complexity. Finally, other contributions~\cite{moulin_universal_2008, amiri_high_2009} addressed the problem of characterizing  probabilistic codes in terms of achievable capacity from an information-theoretical point of view. The latter results have led to several improvements of the decoding functions, see~\cite{meerwald_towards_2011, furon_em_2009}. 

In this paper, we are concerned with ``simple" decoders such as the original one proposed by Gabor Tardos which provides a theoretical proof of performance under threshold-based decisions. Such decoders have a decoding complexity scaling as $O(n)$. Joint decoders, requiring to analyze subsets of (up to) $c$  possible traitors among $n$ users lead to better performance \cite{amiri_high_2009} but for higher complexity. This type of decoder will therefore be out of the scope of this paper.


The original ``simple" Tardos decoder is known to be suboptimal if the collusion channel, \ie the coalition strategy and the coalition size, is known at the decoding side. However as it is unknown in practice, an approach given in \cite{furon_em_2009} solved this problem with an estimated collusion channel. Even though better effectiveness is achieved for small coalition, this approach remains complex for large coalitions. Additionnaly the bound on the false alarm probability is not ensured as the authors of \cite{furon_em_2009} have not solved the threshold based decision issue.

As the original Tardos decoder, our decoder is ``agnostic" because it does not need to estimate the collusion channel. Our decoder addresses the traitor tracing problem under the marking assumption and solves a test under a specific Maximum a Posteriori (MAP) decision rule. In particular, the decision rule is devised under the assumption that the densities of probability of both the traitors strategy and the coalition size follow a \textit{non-informative} law.

We compare our decoder with the symmetric version of the Tardos decoder given in \cite{skoric_tardos_2008}. 
Generating the receiver operating characteristic (ROC) by Monte Carlo simulation, we show that  better result can be obtained with the proposed methodology in all the considered settings. 
In particular, the efficiency of our proposal is presented for common collusion channels presented in the literature and different code lengths. The improvement of the performance is however at the cost of a small increase of the decoding complexity. Indeed the complexity of our decoder scales as $\mathcal{O}(nc)$ and, unlike Tardos decoding, varies therefore linearly with the maximum number of colluders.

The rest of the paper is organized as follows: Section 2 provides the probabilistic model of the problem. Section 3 presents the rationale of our decoding approach and the details of the proposed decoder. Section 4 is concerned with the experimental evaluation. Finally Section 5 gives some concluding remarks.

\section{Notations}
Throughout the paper, we will use the following notations. We use uppercase letters for random variables, lowercase letters for their individual values, and boldface fonts for sequences (or vectors).
$\mathbb{P}_X(x)$ will denote the probability of random variable $X$ evaluated at $x$. However, when there is no possible ambiguity, we often use the shorthand notation: $\mathbb{P}(x)\triangleq\mathbb{P}_X(x)$. The binomial coefficient indexed by $n$ and $k$ is denoted ${n \choose k}$.
\section{Probabilistic Model}
\label{PM}

Let $\mathbf{X}$ $\in \lbrace0,1\rbrace^{m\times n}$ define a length-$m$ binary code for $n$ users. In practical systems, a different column of the code $\mathbf{X}$ denoted $\mathbf{x_j}$ is hidden in the multimedia content delivered to each user $j$. We assume that $c$ users (referred to as the colluders) combine their contents to form a new sequence $\mathbf{y}$ $\in \lbrace0,1\rbrace^m$. In the sequel, we will identify the users participating to the collusion by a vector $\mathbf{s}$ $\in \lbrace0,1\rbrace^n$ defined as follows: $s_j = 1$ if the $j$th user is a colluder and $s_j = 0$ otherwise. Clearly, we have the following relation between $c$ and $\mathbf{s}$:
\begin{align}
c = \sum_i s_i.\nonumber
\end{align}

For a given size of collusion $c$, we assume that all the repartition of the colluders within the users are equally likely, that is
\begin{align}
\mathbb{P}(\mathbf{s}\vert c ) = 1 / {n \choose c}.
\label{eq:model0}
\end{align}

Moreover, it is commonly assumed that the $i$th element of $\mathbf{y}$ only depends on the number of  1's appearing in the colluder's codewords at position $i$. More formally, let $\mathbf{t}\in\{0,\ldots,c\}^{m}$ be a vector whose $i$th element  is the number of symbols ``1" in the colluder sequences at position $i$, $1\leq i\leq m$.  We have therefore
\begin{align}
\mathbf{t}=\mathbf{X}\mathbf{s}. \label{eq:model1}
\end{align}
Given $\mathbf{t}$, the probability of the sequence $\mathbf{y}$ generated by the colluders is totally characterized by the following conditional probability
\begin{align}
\mathbb{P} (\mathbf{y} \vert \mathbf{t}, \mathbf{G})=\prod_i \mathbb{P}(y_i \vert t_i, \mathbf{G}), \label{eq:model2}
\end{align}
where 
\begin{align}
\mathbb{P}_{Y_i\vert T_i, G} (y_i \vert t_i=k,\mathbf{G}) \sim Ber(g_{ki}). \nonumber
\end{align}
Hence, the choice of the matrix $\mathbf{G}$ of Bernoulli parameters fully characterizes the collusion strategy of a coalition of size $c$. For clarity, we do not specify the parameter $c$ in the notation of the matrix $\mathbf{G}$ of size $m\times c$. The elements of $\mathbf{G}$ can be arbitrary except for the elements $g_{ik}$ with $k\in\{0,c\}$ which should obey the so-called ``marking assumption" \cite{boneh_collusion-secure_1998}, that is $g_{i0}=0$ and $g_{ic}=1$ $\forall\, i$.

In practice, the ability of any system to identify the colluders (\ie the vector $\mathbf{s}$) from the observed sequence $\mathbf{y}$ strongly depends on the code $\mathbf{X}$ used to protect the  content. In his seminal paper \cite{tardos_optimal_2008}, Tardos proposed to construct the code in a probabilistic manner as follows:
\begin{equation}
\mathbb{P}(\mathbf{X}\vert\mathbf{p}) = \prod_{i=1}^{m}\prod_{j=1}^{n} \mathbb{P}(x_{ij}\vert p_i),\label{eq:model3}
\end{equation}
where
\begin{align}
\mathbb{P}_{X_{ij}\vert P_{i}}(x_{ij}\vert p_i)&\sim Ber(p_i),\nonumber
\end{align}
and $\mathbf{p}$ denotes the secret vector collecting the Bernoulli parameters $p_i$. Moreover, Tardos proposed a specific distribution to generate the latter parameters:
\begin{align}
\mathbb{P}(\mathbf{p})=\prod_{i=1}^m\mathbb{P}(p_i),\label{eq:model4}
\end{align}
with\footnote{We omit here the cutoff parameter for the sake of simplicity.}
\begin{align}
\mathbb{P}_{P_i}(p_i)&\sim (1/(\pi\sqrt{p_i(1-p_i)})),~\mbox{with~~} p_i \in ]0,1[.\nonumber
\end{align}

In conclusion, we have that the joint probability of the different quantities entering into play in the conception of the observed sequence $\mathbf{y}$ by the colluders defined in $\mathbf{s}$ can be expressed as follows:
\begin{align}
\mathbb{P}(\mathbf{y},\mathbf{t}, \mathbf{X},\mathbf{p}, \mathbf{s}\vert c, \mathbf{G})
=& \mathbb{P}(\mathbf{y}\vert \mathbf{t},\mathbf{G}) \mathbb{P}(\mathbf{t}\vert \mathbf{X},  \mathbf{s}) \mathbb{P}(\mathbf{s}\vert c ) \mathbb{P}(\mathbf{X}\vert \mathbf{p} )\mathbb{P}(\mathbf{p} ),\label{eq:factorisation_joint}
\end{align}
where the different conditional probabilities appearing in the right-hand side of \eqref{eq:factorisation_joint} have been defined in \eqref{eq:model0}, \eqref{eq:model1}, \eqref{eq:model2}, \eqref{eq:model3} and \eqref{eq:model4}. In the next section, we will exploit this sound probabilistic characterization of the system to derive a new colluder detector.

\section{Decoding Description}
\label{MAP}

The ultimate goal of any fingerprinting system is to accurately identify the users responsible of the release of the pirated content. More formally, this requires to properly estimate the ``accusation" vector $\mathbf{s}$ from the observed sequence $\mathbf{y}$. In practice, it is sufficient to accuse at least one guilty user while innocent users are deemed guilty with sufficiently low probability. This task often results in a compromise between accuracy and computational complexity. This section is dedicated to the derivation of a novel decoder offering a good trade-off between these two contradicting goals. In section \ref{sub_con}, we first replace our contribution in the existing literature. Then, in section \ref{sub_map}, we derive the accusation functions defining our decoder.

\subsection{Connections with previous contributions}
\label{sub_con}

This section is dedicated to linking our approach with existing ``simple" decoders. The identification of such decoders is related to a score $\sigma_j$ which gives sufficient information about the involvement of a user $j$ in the forgery of $\mathbf{y}$. The sources of information available at the decoder for the evaluation of one user's score are the forgery $\mathbf{y}$, the sequence of the user $\mathbf{x_j}$ and the secret vector $\mathbf{p}$. In such a context, in order to prevent accusation of innocent users, two scenarios are possible: either all users with a score above a threshold are accused or only the user with the biggest score above the threshold is accused. The decoder is evaluated in terms of soundness and completeness. The decoder is said to be $\epsilon_1$-sound if the false alarm probability is bounded by $\epsilon_1$, and said to be $\epsilon_2$-complete if the false negative probability is bounded by $\epsilon_2$ for a maximum coalition size.

Two kinds of simple decoders exist. The first ones adapt their scores computation to the collusion channel as in \cite{furon_em_2009}. The worst case attacks are still unknown for such decoders and the false alarm probability is not bounded for any coalition sizes. Their effectiveness is experimentally assessed. 
On the contrary, the second class of decoders is independent of the collusion channel as in \cite{tardos_optimal_2008}. Theoretical proofs of soundness and completeness are given in \cite{tardos_optimal_2008} by using Chernoff bounds.  

 

In an informed setup, where the decoder knows the collusion channel and the size of the collusion, the Neyman-Pearson theorem tells us that the optimal discriminative score, say $\sigma_j^{\mathrm{NP}}$, to test whether user $j$ pertains to the collusion or not is as follows:
\begin{align}
\sigma_j^{\mathrm{NP}} =  \frac{\mathbb{P}(\mathbf{y} \vert \mathbf{x}_j, s_j=1, \mathbf{G}, \mathbf{p},c)}{ \mathbb{P}(\mathbf{y} \vert \mathbf{x}_j, s_j=0, \mathbf{G},\mathbf{p},c)}.
\label{eq:informed}
\end{align} 
However this scoring is out of reach since the collusion channel is unknown to the decoder. Some class of agnostic decoders exist where scoring functions are independent of the collusion size $c$ and the collusion strategy $\mathbf{G}$. In~\cite{skoric_symmetric_2008}, the authors proposed a symmetric version of the original Tardos approach. This decoder computes a score for each user as 
\begin{align}
\sigma_j^t = \sum_{i=1}^{m}U(y_i,x_{ij},p_i),
\label{eq:sym_tardos}
\end{align}
with
\begin{align}	
U(1,1,p_i) = \sqrt{(1-p_i)/p_i}, && U(0,0,p_i) = \sqrt{p_i/(1-p_i)} \nonumber
\end{align}
and 
\begin{align}	
U(1,0,p_i) = - U(1,1,p_i), && U(0,1,p_i) = -U(0,0,p_i). \nonumber
\end{align}

This scoring function ensures some kind of separation between the distributions of scores of the innocent and traitor users  for any collusion channel compliant with the marking assumption. It then permits to derive an appropriate threshold which guarantees to bound the false alarm probability for a given code length $m$.  

However the authors of \cite{furon_em_2009} have shown the huge gap between the symmetric Tardos decoder and the informed decoder of equation \eqref{eq:informed}. Hence, as the collusion channel is unknown in practice, they have proposed to estimate it. Their approach is based on the so-called Expectation-Maximization algorithm. The authors assume that the collusion strategy is constant for all positions $i$ for the sake of  simplifying the mathematical model.
Our assumption on the collusion channel is more general in the sense that our decoder considers all possible strategies at each $i$th position. 


\subsection{MAP Decoding with Non-informative Priors}
\label{sub_map}

The challenge of robust and effective detection procedures stands in the fact that some parameters (namely $\mathbf{G}$ and $c$) of the model are actually unknown to the decoder. Now, a very common approach in Bayesian statistics consists in defining non-informative priors on the unknown quantities and marginalize them out from the joint probability characterizing the system. By ``non-informative" prior, it is usually understood a probability distribution not favoring any of the possible realizations of the considered random variable. 

More specifically, our approach consists in exploiting the following joint probability to derive our decoder:
\begin{align}
\mathbb{P}(\mathbf{y},\mathbf{t}, \mathbf{X},\mathbf{p},\mathbf{s})
=& \sum_c\left(\int \mathbb{P}(\mathbf{y},\mathbf{t}, \mathbf{X},\mathbf{p},\mathbf{s}\vert c,\mathbf{G})\mathbb{P}(\mathbf{G}) d\mathbf{G}\right)\mathbb{P}(c),  \nonumber
\end{align}
where $\mathbb{P}(\mathbf{y},\mathbf{t}, \mathbf{X},\mathbf{p}, \mathbf{s}\vert c, \mathbf{G})$ has been specified in~\eqref{eq:factorisation_joint} and $\mathbb{P}(c)$,  $\mathbb{P}(\mathbf{G})$
are non-informative priors which will be defined hereafter. In turn, these joint probabilities can be marginalized to compute the following likelihood ratio
\begin{align}
\sigma_j^{\mathrm{MAP}}\triangleq \frac{\mathbb{P}(s_j=1\vert \mathbf{y},\mathbf{x}_{j},\mathbf{p})}{\mathbb{P}(s_j=0\vert\mathbf{y},\mathbf{x}_{j},\mathbf{p})}=\frac{\mathbb{P}(\mathbf{y},\mathbf{x}_{j},\mathbf{p}, s_j=1)}{\mathbb{P}(\mathbf{y},\mathbf{x}_{j},\mathbf{p}, s_j=0)}.\nonumber
\end{align}

At the decoding side, for a given code length, it is illusive to chase more than say $c_{\max}$ colluders.
The size of the collusion is seen as a discrete random variable ranging from 1 to $c_{\max}$.
Therefore, its non-informative prior distribution is the uniform law, that is 
\begin{equation}
\mathbb{P}(c)=\frac{1}{c_{\max}}.\label{eq:pcuniform}
\end{equation}
As for the collusion channels, we first assume the statistical independence of the parameters:
\begin{equation}
\mathbb{P}(\mathbf{G})=\prod_{i,k}\mathbb{P}(g_{ik}).\nonumber
\end{equation}

We then enforce the marking assumption: $g_{i0}=1-g_{ic}=0,\,\forall i$. The other parameters $G_{ik}$ are seen as continuous random variables ranging in $[0,1]$. Not favoring any of the possible realizations of $G_{ik}$, we set the intuitive uniform law as a non-informative prior distribution. Then marginalizing over $g_{ik}$ for $ k \in \lbrace1,...,c-1\rbrace$ leads to:
\begin{align}
\mathbb{P}(y_i \vert t_i=k ) =& \int_{0}^{1} \mathbb{P}(g_{ik}) g_{ik}^y (1-g_{ik})^{1-y} d_{g_{ik}}\\
 =& \int_{0}^{1} g_{ik}^y (1-g_{ik})^{1-y} d_{g_{ik}}=1/2.
\label{eq:uniform}
\end{align}

Notice that all Beta law Beta($\alpha$,$\beta$) of equal parameters, such as the well known Jeffreys prior of parameters $\alpha=\beta=1/2$, give an equivalent result as in \eqref{eq:uniform}. Indeed if
\begin{equation}
\label{eq_theta}
P(g_{ik}) = \frac{g_{ik}^{\alpha-1} (1-g_{ik})^{\beta-1}}{B(\alpha,\beta)},
\end{equation}
with
\begin{equation}
B(\alpha,\beta) = \int_0^1 v^{\alpha-1}(1-v)^{\beta-1}dv\\  \nonumber
\end{equation}
and if $\alpha=\beta$,
\begin{equation}
\mathbb{P}(y_i \vert t_i=k )=1/2.
\label{eq_beta}
\end{equation}

The uniform law is then just a Beta law of parameters $\alpha=\beta=1$ in \eqref{eq_theta}. 

Let us note that if the realizations of $c$ and $\mathbf{G}$ obeyed probabilities $\mathbb{P}(c)$ and $\mathbb{P}(\mathbf{G})$, $c$- and $\mathbf{G}$-blind optimal Neyman-Pearson test would result in a simple thresholding of $\sigma_j^{\mathrm{MAP}}$.

Let us then particularize the expression of $\sigma_j^{\mathrm{MAP}}$ to the particular hypotheses introduced in \eqref{eq:pcuniform} and \eqref{eq_beta}. We have
\begin{equation}
\label{eq_test}
\sigma_j^{\mathrm{MAP}}=\frac{\sum_{c=1}^{c_{max}} \mathbb{P}(\mathbf{y} \vert s_j=1,\mathbf{x_j},c) \mathbb{P}(s_j=1 \vert c)}{\sum_{c=1}^{c_{max}} \mathbb{P}(\mathbf{y} \vert s_j=0,\mathbf{x_j},c) \mathbb{P}(s_j=0 \vert c)},
\end{equation}
where
\begin{eqnarray}
\mathbb{P}(s_j=1 \vert c ) &=& c/n,\nonumber \\
\mathbb{P}(s_j=0 \vert c ) &=& (n-c)/n,\nonumber
\end{eqnarray}
%
\begin{align}
\mathbb{P}(\mathbf{y} \vert s_j,\mathbf{x}_{j},c) 
=& \prod_i \mathbb{P}(y_i \vert s_j,x_{ij},c) \nonumber\\
=& \prod_i\sum_{t_{i}=0}^{c} \mathbb{P}(y_i,t_i \vert s_j,x_{ij})\nonumber\\
=&\prod_i\sum_{t_{i}=0}^{c} \mathbb{P}(y_i \vert t_i) \mathbb{P}(t_i \vert s_j, x_{ij},c)\nonumber
\end{align}
and
\begin{align}
   \mathbb{P}(t_i \vert s_j=1, x_{ij},c)&={c-1 \choose t_i-x_{ij}} p_i^{t_i-x_{ij}} (1-p_i)^{c-1-t_i+x_{ij}}, \nonumber\\
   \mathbb{P}(t_i \vert s_j=0,x_{ij},c)&={c \choose t_i}p_i^{t_i} (1-p_i)^{c-t_i}. \nonumber
\end{align}%
Particularizing these expressions to \eqref{eq_beta}, we obtain after some algebraic manipulations:

\begin{eqnarray}
\mathbb{P}(y_i \vert s_j=1,x_{ij})&=& 1/2 \times (1+(-1)^{y_{i}}((1-x_{ij})\nonumber\\
&\times & (1-p_i)^{c-1} - x_{ij}p_i^{c-1})),\nonumber\\
\mathbb{P}(y_i \vert s_j=0,x_{ij})&=& 1/2 \times (1+(-1)^{y_i}((1-p_i)^{c}-p_i^{c})).\nonumber
\end{eqnarray}

\subsection{Numerical solution}
The evaluation of the large products appearing in equation~\eqref{eq_test} suffers from numerical problems w.r.t.
machine finite precision.
The logarithm translates products into sums.
However taking the logarithm of our test does not give a simple formulation due to the sum over the possible coalition sizes.
We resort to generalized maximum function $M_{g}$ as shown in~\cite{robertson_comparison_1995}:
\begin{eqnarray}
M_{g}(a,b)&\triangleq&\log(\exp(a) + \exp(b)) \nonumber\\
 &=& \max(a,b) + \log(1 + e^{- \vert a - b \vert}),\nonumber\\
M_{g}(a,b,c)&\triangleq&\log(\exp(a) + \exp(b) + \exp(c))\nonumber\\
 &=& M_g\left(M_g(a,b),c\right).\nonumber
\end{eqnarray}
The test~\eqref{eq_test} can be formulated as follows in the logarithmic domain:
\begin{equation}
\sigma_j^{\mathrm{MAP}}=\log\left(\sum_{c=2}^{c_{max}}e^{A1_{c}}\right) - \log\left(\sum_{c=2}^{c_{max}}e^{A2_{c}}\right), \nonumber
\end{equation}
with
\begin{eqnarray}
A1_c &=& \log\frac{c}{n}+\sum_{i=1}^{m}{\log\mathbb{P}(y_i \vert s_j=1,x_{ij},c)}, \nonumber\\
A2_c &=& \log\frac{n-c}{n}+\sum_{i=1}^{m}{\log\mathbb{P}(y_i \vert s_j=0,x_{ij},c)}. \nonumber
\end{eqnarray}
The generalized maximum function gives the following recursive expression of the test:
\begin{equation}
\sigma_j^{\mathrm{MAP}} =  M_g(M_g(...), A1_{c_{max}}) - M_g(M_g(...), A2_{c_{max}}). \nonumber
\end{equation}

\section{Simulation Results}
The experimental investigation 
is composed of two parts.
The first one presents
the effectiveness of the method for different collusion channels. The second part presents
the effectiveness of the method for different code lengths. We used the classical Monte Carlo estimator to estimate the performance. 

We compared our approach with two decoders. The first decoder is the symmetric version of the Tardos decoder given in~\eqref{eq:sym_tardos}. The second decoder is the informed decoder given in~\eqref{eq:informed}. 

Receiver Operating Characteristics (ROC) curves plots are used to compare the three decoders. Deterministic or \textit{random} strategies are considered as follows. ``Minority" and ``Majority" are deterministic strategies where the less or the most frequent symbol is  put in the pirated sequence. ``Uniform", ``Coin flip" and `Worst case attack" are random strategies. In the ``Uniform" strategy, the colluders uniformly-randomly choose
one of their symbols. In the ``Coin flip" strategy, the colluders flip a fair coin to choose a symbol. 
Finally in the ``Worst case attack" (wca) strategy, the colluders minimize the mutual information between the symbols of the pirated copy and each of their codewords. It is considered to be the worst attack against the best achievable simple decoder from an information-theoretical viewpoint. This strategy is obtained by a minimization algorithm described in \cite{furon_worst_2009}. This minimization gives a stationary attack channel as introduced in section~\ref{sub_con}. 

In \cite{tardos_optimal_2008}, the efficiency of the code in term of error bound is proved over all random choices of the code, \ie for all random choices of the secret vector $\mathbf{p}$, the dictionary of users $\mathbf{X}$ and the strategy $\mathbf{G}$. In our experiment, for one realization of the Monte Carlo, all these variables are randomly sampled.
For one realization we use $n=10^3$ users, with $c$ fixed before running the experiment. We then compute the scores with three decoders. At each realization we store $2\times 3$ scores, as we keep only the biggest score of the colluders and the biggest score of the innocents. Each run encompasses $10^4$ realizations.

The false positive is related to $\sigma_{x_{inn}}$, the vector of all innocent-user scores as 
\begin{align}
pfa = \mathbb{P}(max(\sigma_{x_{inn}})\geq \tau).\nonumber 
\end{align}
Hence a false alarm event occurs when at least one innocent user is accused, \ie one innocent-user score is above the considered threshold $\tau$.

The false negative is related to $\sigma_{x_{coll}}$, the vector of all colluder scores, as
\begin{align}
pfn = \mathbb{P}(max(\sigma_{x_{coll}})<\tau).\nonumber
\end{align} 
The false negative event occurs when all colluders are missed, \ie when all colluder scores are below the considered threshold $\tau$.

\subsection{Stability over different collusion channels}
\label{subsec_strat}
%
%
\begin{figure}
\centering
\includegraphics[scale=0.70]{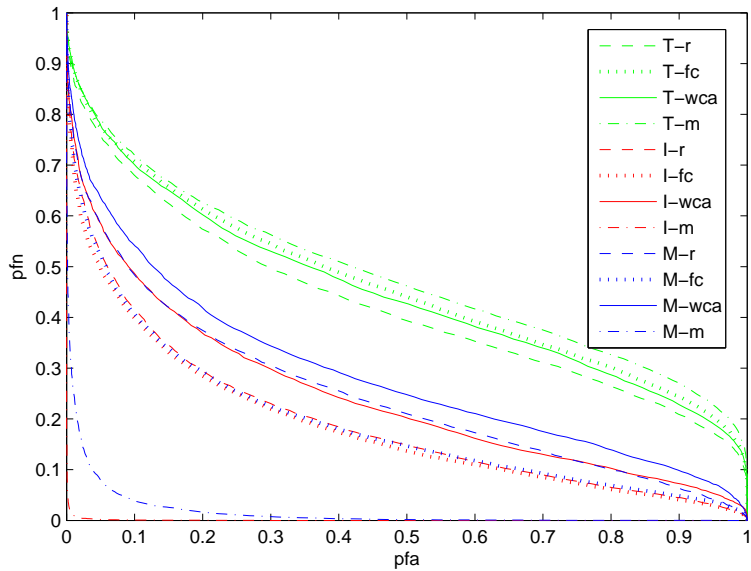}
\caption{ROC curves of the informed decoder, the MAP blind decoder and the symmetric Tardos decoder for $4$ different collusion channels with $m=300$, $c=6$ and $n=1000$ users.}
\label{fig_strat}
\end{figure}	
Figure~1 shows the ROC curves for the symmetric Tardos decoder, the informed decoder and our MAP blind decoder. We consider a fingerprinting code of length $m=300$, a maximum number of colluders $c_{max}=10$ and a true number of colluders $c=6$. The legend of Figure~1 is set as follows. The uppercase letters ``T", ``I", ``M" are used in this order for the \textbf{T}ardos decoder, the \textbf{I}nformed decoder and the \textbf{M}AP blind decoder. The lowercase letters ``r",``fc",``wca",``m" states for  \textbf{r}andom, \textbf{f}lip \textbf{c}oin, \textbf{w}orst-\textbf{c}ase \textbf{a}ttack and \textbf{m}inority strategies.

For all strategies our decoder leads to enhanced performance as compared to Tardos decoder. Not surprisingly, the informed decoder gives the best performance for all strategies. The stability of the Tardos decoder performance is shown for this set of collusion strategies: unlike the informed and MAP blind  decoders, its performance does not vary a lot with the considered strategy. For some strategies, this stability  is however achieved at the expense of a loss of performance with respect  to the informed and the MAP blind decoders. In particular, the largest gap between the performance of Tardos and informed/MAP blind decoders is reached for the minimum strategy which appears to be the more damaging for Tardos approach \cite{furon_em_2009}.  
The wca strategy is the worst strategy against the Informed decoder. It is also the worst strategy against our MAP blind decoder for this set of strategies. However it is important to mention that the wca strategy has not been proved to be the worst attack against the MAP blind decoder.	

Even if the MAP blind decoder is not quite as good as the informed decoder for the flip coin strategies, its performance is very closed to the informed decoder, the ROC curves are almost overlapped in this case. This is consistent with the coin flip strategy, because only the true coalition size is unknown for the MAP blind decoder. Notice that the performance of the MAP blind decoder against the flip coin attack is a little bit better than the performance of the informed decoder for the random strategy. Nevertheless, these last three considered configurations lead almost to the same performance. It is also the case for the worst case attack against the informed decoder and for the random attack against the MAP Blind decoder.

\subsection{Evaluation over different code lengths}

\begin{figure}
\centering
\includegraphics[scale=0.70]{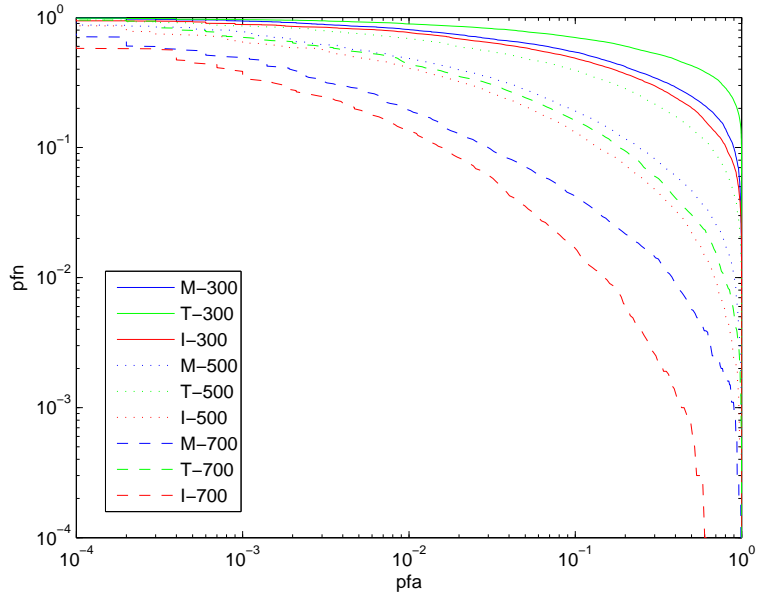}
\caption{ROC curves of the informed decoder, the MAP blind decoder and the symmetric Tardos decoder for $3$ different code lengths and the wca strategy with $c=6$ and $n=1000$ users.}
\label{fig_length}
\end{figure}	
Figure~2 presents the ROC curves for the symmetric Tardos decoder, the informed decoder and our MAP blind decoder for different code lengths. We evaluate the performance of the decoders for the wca attack since it is the worst attack against our MAP blind decoder among the considered strategies. We use the same set of parameters as in Figure~1 with the same legend terminology. The probability of false alarm and the probability of false negative are set in logarithmic scale in Figure~2. 

For all lengths, our decoder results in less decoding errors compared to the symmetric Tardos decoder. In particular, our decoder performance is closer to the informed decoder performance than Tardos decoder. Moreover, the gap between the performance of the Tardos and the MAP blind decoders  increases as the length of the code increases.



\section{Conclusion}
Our blind Maximum A Posteriori approach works for any probabilistic codes under the marking assumption with acceptable complexity. Promising results compared to the symmetric Tardos decoder are presented.  The preliminary results presented here open however some important questions:
\begin{enumerate}
\item What is the behaviour of our decoder if the true coalition size is above the maximum coalition size set to the decoder?
\item Is our decoder better than estimation-based decoders, such as in \cite{furon_em_2009}, against time varying attacks and stationary attacks?
\item How is linked our decoder functions with the Tardos ones? Some preliminary numerical results, not presented here, gives some correlations in particular asymptotic cases.   
\end{enumerate}

These three last issues will be addressed in our future research as the study of the setting of a proper threshold so as to bound false alarm probability.

\section*{Acknowledgment}
The authors thank Teddy Furon for his valuable help in writing this paper.

\bibliographystyle{IEEEtran}
\bibliography{mmsp2013}

\end{document}